\documentclass[aps, prd, reprint, superscriptaddress, nofootinbib]{revtex4-2}

\usepackage{amsmath,amssymb,amsfonts}
\usepackage{slashed}

\usepackage{orcidlink}
\usepackage[dvipsnames,table]{xcolor}

\hypersetup{colorlinks, urlcolor=BlueViolet, citecolor=Plum, linkcolor=PineGreen}

\newcommand{\braket}[1]{\langle #1 \rangle}
\newcommand{\bra}[1]{\langle #1 |}
\newcommand{\ket}[1]{| #1 \rangle}

\begin{document}

\title{Spin-independent scattering of pseudoscalar-mediated dark matter}

\author{Nicole F. Bell
\orcidlink{0000-0002-5805-9828}}
\affiliation{ARC Centre of Excellence for Dark Matter Particle Physics, School of Physics, The University of Melbourne, Victoria 3010, Australia}

\author{Giorgio Busoni
\orcidlink{0000-0002-8527-0768}}
\affiliation{ARC Centre of Excellence for Dark Matter Particle Physics and CSSM, Department of Physics, Adelaide University, South Australia 5005, Australia}

\author{Peter Cox
\orcidlink{0000-0002-6157-3430}}
\affiliation{ARC Centre of Excellence for Dark Matter Particle Physics, School of Physics, The University of Melbourne, Victoria 3010, Australia}

\author{Laura W. Fang
\orcidlink{0009-0003-6602-0792}}
\email{laura.fang@student.unimelb.edu.au}
\affiliation{ARC Centre of Excellence for Dark Matter Particle Physics, School of Physics, The University of Melbourne, Victoria 3010, Australia}

\author{John Gargalionis
\orcidlink{0000-0002-0745-7121}}
\affiliation{ARC Centre of Excellence for Dark Matter Particle Physics and CSSM, Department of Physics, Adelaide University, South Australia 5005, Australia}

\author{Jayden L. Newstead
\orcidlink{0000-0002-8704-3550}}
\affiliation{ARC Centre of Excellence for Dark Matter Particle Physics, School of Physics, The University of Melbourne, Victoria 3010, Australia}

\author{Ewan N. V. Wallace
\orcidlink{0009-0000-9092-8029}}
\email{ewan.n.wallace@student.unimelb.edu.au}
\affiliation{ARC Centre of Excellence for Dark Matter Particle Physics, School of Physics, The University of Melbourne, Victoria 3010, Australia}
\affiliation{ARC Centre of Excellence for Dark Matter Particle Physics and CSSM, Department of Physics, Adelaide University, South Australia 5005, Australia}

\author{Martin J. White
\orcidlink{0000-0001-5474-4580}}
\affiliation{ARC Centre of Excellence for Dark Matter Particle Physics and CSSM, Department of Physics, Adelaide University, South Australia 5005, Australia}

\author{Anthony G. Williams
\orcidlink{0000-0002-1472-1592}}
\affiliation{ARC Centre of Excellence for Dark Matter Particle Physics and CSSM, Department of Physics, Adelaide University, South Australia 5005, Australia}

\begin{abstract}
Dark matter with pseudoscalar couplings provides a well-motivated scenario in which direct-detection signals are suppressed at tree level, since the scattering off nuclei is both spin-dependent and momentum suppressed.  While spin-independent scattering is absent at tree level, it arises at one loop and can provide the leading direct-detection signal. We revisit this scenario in a general sub-electroweak effective field theory with a light pseudoscalar mediator, including interactions through to mass-dimension-six. We compute the matching onto the quark and gluon operators relevant for direct detection to determine whether this scenario could be detectable at future experiments, while also requiring consistency with the observed dark matter relic-abundance and indirect-detection limits. We find that while models with a pseudoscalar mediator can generate spin-independent cross sections above the neutrino floor, this generally requires additional new physics below the TeV scale.

\end{abstract}

\maketitle


\section{Introduction}
\label{sec:introduction}

Direct-detection experiments can search for dark matter (DM) through the nuclear recoils produced when DM scatters with nuclei in the detector. These experiments are most sensitive to interactions that generate unsuppressed spin-independent (SI) scattering, since such interactions are coherently enhanced for large nuclei. The increasingly strong null results from direct-detection experiments therefore place severe constraints on models in which SI scattering arises at tree level~\cite{XENON:2025vwd, LZ:2024zvo, PandaX:2024qfu}.

This motivates the study of scenarios in which the leading direct-detection signal is absent or suppressed. In the non-relativistic description of DM--nucleon scattering, different high-energy interactions map onto different nuclear responses, some of which are spin-dependent and/or suppressed by powers of the small momentum transfer or DM velocity~\cite{Fan:2010gt,Fitzpatrick:2012ix,Dent:2015zpa,Bishara:2017pfq,Hisano:2015bma}. Pseudoscalar-mediated interactions provide a particularly important example. For DM that is coupled to Standard Model (SM) matter through pseudoscalar interactions, the tree-level elastic scattering amplitude is spin-dependent and momentum suppressed. This spin-dependent direct-detection signal can be extremely small, even when the same interactions allow for efficient annihilation in the early Universe~\cite{Ipek:2014gua,Dolan:2014ska,Berlin:2015wwa,Bauer:2017ota,Fan:2015sza}.

The absence of an appreciable tree-level interaction does not, however, imply that models with pseudoscalar mediators are invisible to direct detection. In the non-relativistic regime, kinematic suppressions can be more important than loop factors: a loop-induced SI amplitude may dominate over a tree-level amplitude that is spin-dependent and momentum suppressed~\cite{Freytsis:2010ne,Haisch:2013uaa,Bell:2018zra}. This is well known in the literature and has been studied within the context of specific models with a pseudoscalar mediator~\cite{Arcadi:2017wqi,Li:2018qip,Abe:2018emu,Azevedo:2018exj,Ghorbani:2018pjh,Ertas:2019dew,Arcadi:2022lpp,Li:2019fnn,Beenakker:2025mhf,Alanne:2020xcb}.

The aim of this paper is to determine, more generally, when loop effects can generate observable direct-detection signals in DM models with pseudoscalar interactions. Rather than a particular simplified model or ultraviolet scenario, we therefore consider a general low-energy EFT that allows us to draw broader, model-independent conclusions.

The EFT that we consider is defined by a small number of assumptions. The new degrees of freedom are the DM (which we take to be a Dirac fermion) and a light pseudoscalar mediator that couples to quarks and/or gluons. The quark couplings are assumed to obey minimal flavour violation, so that the new interactions inherit the flavour structure of the SM and avoid the most stringent flavour constraints~\cite{DAmbrosio:2002vsn,Dolan:2014ska}. Finally, we include operators through mass dimension six that are compatible with a purely pseudoscalar portal between DM and the SM at tree level. The central question is then: in this model-agnostic setting, can loop-induced SI scattering produce an experimentally relevant direct-detection signal while remaining compatible with the relic density constraint?

We answer this question by computing the one-loop matching of the EFT onto the scalar and twist-2 quark and gluon operators that generate SI DM--nucleon scattering~\cite{Hill:2014yxa,DEramo:2014nmf,Hisano:2015bma,Bishara:2017pfq,Brod:2017bsw}. Our results show that after imposing the relic density requirement, loop-induced SI scattering can reach experimentally relevant rates, but only in certain cases. In particular, this requires the new physics of the UV completion to lie below the TeV scale. 

The remainder of this paper is organised as follows. In Sec.~\ref{sec:eft-setup}, we introduce the sub-electroweak EFT. In Sec.~\ref{sec:SI-scattering}, we compute the loop-level matching onto the quark and gluon operators relevant for SI direct detection. In Sec.~\ref{sec:constraints}, we discuss the DM annihilation channels, and combine these ingredients in Sec.~\ref{sec:results} to determine the parameter space that may be probed via direct detection. We conclude in Sec.~\ref{sec:conclusion}.


\section{Pseudoscalar Mediator EFT}
\label{sec:eft-setup}

This section defines the EFT that we consider throughout the paper. We work below the electroweak scale and retain both the DM $\chi$ and the mediator $a$ as low-energy degrees of freedom. Our rationale for retaining the mediator in the effective theory is given in App.~\ref{app:contact}.

Our setup is defined by the following assumptions. We take the DM field $\chi$ to be a Dirac fermion and the mediator $a$ to be a real pseudoscalar, both singlets under the SM gauge group. We impose a $\mathbb{Z}_2$ symmetry under which $\chi$ is odd and all other fields are even, thereby ensuring the stability of the DM. We further assume that all interactions between $\chi$ and the SM are mediated by $a$. For the interactions with quarks, we impose minimal flavour violation (MFV), such that the couplings are flavour diagonal and proportional to the corresponding quark masses. This preserves the flavour structure of the SM and avoids the stringent constraints from flavour-changing neutral currents~\cite{DAmbrosio:2002vsn,Dolan:2014ska}. We also assume that there is no new $CP$-violation and take the Wilson coefficients to be real. 

The pertinent Lagrangian is
\begin{equation}
\begin{split}
    \label{eq:simpl-model-full}
    -\mathcal{L}&\supset C_{\chi}a\bar{\chi}i\gamma^5\chi+\frac{C_{aq}}{\Lambda}\sum_{q \ne t}m_q a\bar{q}i\gamma^5 q \\
    &+\frac{C_{a\tilde{G}}}{\Lambda}\frac{\alpha_s}{4\pi}aG_{\mu\nu}^{A}\tilde{G}^{A\mu\nu}+\frac{C_{a^2q}}{\Lambda^2}\sum_{q \ne t}m_q a^2\bar{q}q  \\
    &+\frac{C_{a^2G}}{\Lambda^2}\frac{\alpha_s}{4\pi}a^2 G_{\mu\nu}^A G^{A\mu\nu} \ ,
\end{split}
\end{equation}
where $\Lambda$ denotes the scale of the underlying UV completion. We include operators up to mass-dimension-six that couple the pseudoscalar to quark and gluon bilinears, but neglect the operator $a^3 \bar{q}i\gamma^5 q$, which is not relevant for direct detection at one loop. The operators $(\partial_\mu a)\bar\chi\gamma^\mu\chi$ and $(\partial_\mu a)\bar\chi\gamma^\mu\gamma^5\chi$ are not included in Eq.~\eqref{eq:simpl-model-full} as they may be removed by application of the equations of motion. We have defined the gluon operators $aG\tilde{G}$ and $a^2GG$ with a factor of $\alpha_s/4\pi$, since these operators generically arise at loop level in UV completions. Note that even in the absence of additional exotic degrees of freedom at high energies, integrating out the top quark generates one-loop contributions to both of these operators if the pseudoscalar also couples to the top quark. The renormalisation-group (RG) evolution and mixing of the Wilson coefficients in Eq.~\eqref{eq:simpl-model-full} are discussed in App.~\ref{app:running}.


\section{Loop-induced SI scattering}
\label{sec:SI-scattering}

In this section, we compute the SI scattering cross section mediated by the pseudoscalar $a$ at loop level. This involves matching the pseudoscalar EFT onto a low-energy basis of DM--parton effective interactions, comprising scalar and twist-2 operators with quarks and gluons. The methodology we adopt for this calculation follows standard treatments in the literature; see, e.g., Refs.~\cite{Hisano:2010ct,Hisano:2015bma}.

We integrate out the pseudoscalar mediator at the scale $\mu=m_a$ to obtain the effective interactions between the DM and quarks/gluons. In the small momentum-transfer limit relevant for direct detection, this leads to an effective Lagrangian of the form
\begin{equation}
\label{eq:DM-quarkglu-effective}
    \mathcal{L}_\mathrm{eff}^\mathrm{SI} = \mathcal{L}_\text{scalar} + \mathcal{L}_\text{twist-2} ,
\end{equation}
where
\begin{equation}
    \mathcal{L}_\text{scalar} = \sum_{q=u,d,s} \mathcal{C}_{q}^{(0)} m_q\bar{\chi}\chi \bar{q}q + \mathcal{C}_{G}^{(0)}\left( -\frac{9\alpha_s}{8\pi}\bar{\chi}\chi G^A_{\mu\nu}G^{A\mu\nu} \right) , \label{eq:Leff-scalar}
\end{equation}
and
\begin{equation}
\begin{split}
    \mathcal{L}_\text{twist-2} &= \sum_{q \ne t} \left(\mathcal{C}_{q}^{(1)}(\bar{\chi}i\partial^\mu\gamma^\nu\chi)\mathcal{O}^q_{\mu\nu}+\mathcal{C}^{(2)}_q(\bar{\chi}i\partial^\mu i\partial^\nu \chi)\mathcal{O}^q_{\mu\nu}\right) \\
    &\quad + \mathcal{C}_{G}^{(1)}(\bar{\chi}i\partial^\mu\gamma^\nu\chi)\mathcal{O}^G_{\mu\nu}+\mathcal{C}_{G}^{(2)}(\bar{\chi}i\partial^\mu i\partial^\nu \chi)\mathcal{O}^G_{\mu\nu}, \label{eq:Leff-twist2}
    \end{split}
\end{equation}
with the twist-2 QCD operators
\begin{align}
    \mathcal{O}^q_{\mu\nu} &\equiv \frac{i}{2}\bar{q}\left(D_\mu\gamma_\nu+D_\nu\gamma_\mu-\frac{1}{2}g_{\mu\nu}\slashed{D}\right)q, \\
    \mathcal{O}^G_{\mu\nu} &\equiv G^{A\sigma}_{\mu} G^{A}_{\sigma\nu}+\frac{1}{4}g_{\mu\nu}G^A_{\alpha\beta}G^{A\alpha\beta}.
\end{align}
Note that although the twist-2 operators in Eq.~\eqref{eq:Leff-twist2} are of higher mass dimension than the scalar operators in Eq.~\eqref{eq:Leff-scalar}, their contributions remain unsuppressed in the non-relativistic limit when $m_\chi > m_a$ and hence they cannot be neglected.\footnote{The appropriate non-relativistic power counting can be made manifest by using heavy DM effective theory (see e.g. Ref.~\cite{Hill:2014yxa}).}

The scalar and twist-2 sectors do not mix under renormalisation group running and, following the approach used in Refs.~\cite{Hisano:2010ct,Hisano:2015bma}, it is convenient to treat the contributions from the two sectors independently. 

In the scalar sector, we retain only the light quark and gluon operators in Eq.~\eqref{eq:Leff-scalar}, integrating out the charm and bottom quarks, in addition to the pseudoscalar mediator. We include contributions to $\mathcal{C}_{q}^{(0)}$ and $\mathcal{C}_{G}^{(0)}$ up to $\mathcal{O}(\alpha_s)$. The scalar operators are RG-invariant to this order in $\alpha_s$ (see e.g.~\cite{Hill:2014yxa}) and we therefore integrate out all fields in a single step, rather than at different matching scales associated with their mass thresholds. This allows us to more straightforwardly compute important two-loop contributions that are discussed in Sec.~\ref{subsec:DMgluon}.

In the twist-2 sector, we instead retain the heavy quark operators in Eq.~\eqref{eq:Leff-twist2}, since the relevant hadronic matrix elements can be expressed in terms of the parton distribution functions and directly evaluated at the scale $\mu=m_a$.

In the following two subsections, we discuss the matching onto the quark and gluon operators, respectively. Since we are interested in the non-relativistic limit, we work to zeroth order in the momentum transfer $q$. Further, given that the momentum $p$ of the parton in the nucleon is typically of order $\Lambda_\text{QCD}$---which is small relative to the range of DM masses we consider---we expand to first order in $p\cdot k/m_\chi^2$ (for DM momentum $k$) and zeroth order in $p^2/m_\chi^2$.

\subsection{DM--quark operators}
\label{subsec:DMquark}

\begin{figure}[t]
    \centering
    \includegraphics{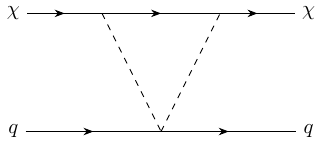}
    \includegraphics{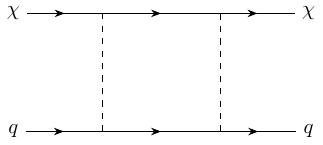}
    \includegraphics{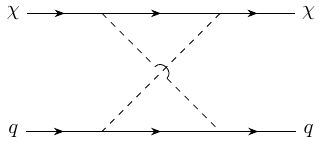}
    
    \caption{One-loop contributions to pseudoscalar-mediated DM-quark scattering. The upper diagram is induced by the $a^2\bar{q}q$ interaction, while the lower two diagrams are induced by the $a\bar{q} \gamma^5 q$ interaction.}
    \label{fig:quarkdiags}
\end{figure}

Contributions to the quark operators, both scalar and twist-2, are generated by the one-loop diagrams in Fig.~\ref{fig:quarkdiags}. We begin by considering the upper diagram, which proceeds via the $a^2\bar{q}q$ interaction in Eq.~\eqref{eq:simpl-model-full}. This generates a contribution to the scalar operator $\bar{\chi}\chi\bar{q}q$, with
\begin{equation} \label{eq:Cq0_a2qq}
    \mathcal{C}^{(0)}_q \supset \frac{C_\chi^2 C_{a^2q}}{\Lambda^2} \frac{m_\chi}{8\pi^2} F_0 ,
\end{equation}
where the loop function $F_0(m_\chi,m_a)$ is given in App.~\ref{app:loopfunctions}.

Next, consider the lower two diagrams in Fig.~\ref{fig:quarkdiags}, which proceed via the $a\bar{q} \gamma^5 q$ interaction. These diagrams generate contributions to the scalar operator as well as the twist-2 DM-quark operators. We find
\begin{align}
    \mathcal{C}^{(0)}_q &\supset \frac{C_\chi^2 C_{aq}^2}{\Lambda^2}\frac{m_q^2 m_\chi}{64\pi^2} (6F_1 + F_2),\\
    \mathcal{C}^{(1)}_q &= \frac{C_\chi^2 C_{aq}^2}{\Lambda^2} \frac{m_q^2}{8\pi^2} F_1,\\
    \mathcal{C}^{(2)}_q &= \frac{C_\chi^2 C_{aq}^2}{\Lambda^2} \frac{m_q^2}{16\pi^2 m_\chi} F_2,
\end{align}
where the loop functions $F_{n}(m_\chi,m_a)$ are given in App.~\ref{app:loopfunctions}. These have been previously calculated in Ref.~\cite{Abe:2018emu}. Note that there is also an additional contribution to a higher-twist quark operator; we ignore this term because it is negligible for direct detection~\cite{Hisano:2017jmz}.

\subsection{DM--gluon operators}
\label{subsec:DMgluon}

\begin{figure}[t]
    \centering
    \includegraphics{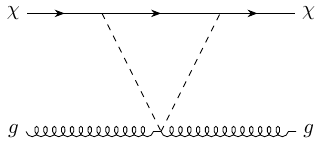}
    \includegraphics{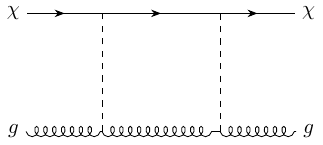}
    \includegraphics{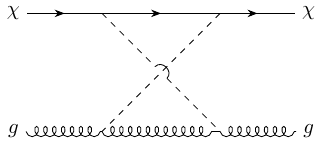}
    
    \caption{One-loop contributions to pseudoscalar-mediated DM-gluon scattering. The upper diagram is induced by the $a^2GG$ interaction, while the lower two diagrams are induced by the $aG\tilde{G}$ interaction.}
    \label{fig:gluondiagrams}
\end{figure}

\begin{figure}[t]
    \centering
    \includegraphics{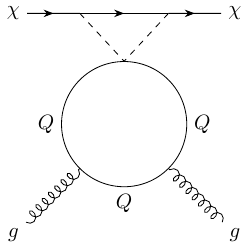}\\
    \includegraphics{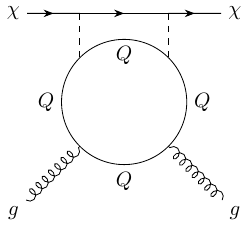}\\
    \includegraphics{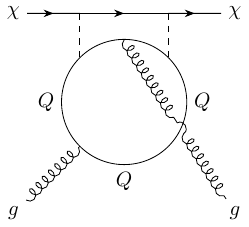}
    
    \caption{Two-loop heavy-quark contributions to the $\bar{\chi}\chi G^A_{\mu \nu}G^{A\mu\nu}$ operator, where $Q$ denotes $\{b,c\}$. The upper diagram is induced by the $a^2\bar{q}q$ coupling and the lower two diagrams by the $a\bar{q} \gamma^5 q$ coupling.}
    \label{fig:twoloop}
\end{figure}

There are two types of contributions to the DM-gluon operators. First, the pseudoscalar interactions with gluons, $aG\tilde{G}$ and $a^2GG$, give rise to the one-loop diagrams in Fig.~\ref{fig:gluondiagrams}. Second, pseudoscalar interactions with the heavy quarks ($q=c,b$) generate the two-loop contributions in Fig.~\ref{fig:twoloop}. 

The importance of two-loop contributions such as these is well known~\cite{Hisano:2010ct,Abe:2018emu}. They contribute to the DM-nucleon cross section at the same order in $\alpha_s$ as the one-loop light quark contributions in Fig.~\ref{fig:quarkdiags}. This is due to the fact that the hadronic matrix elements satisfy $\bra{N}m_q \bar{q}q\ket{N} \sim \bra{N}\frac{\alpha_s}{\pi} G G\ket{N}$ (and is the reason for including the factor of $\alpha_s$ in the normalisation of the gluon operator in Eq.~\eqref{eq:Leff-scalar}).

Note also that the (one-loop) contributions to the gluon operators in Fig.~\ref{fig:gluondiagrams} are in fact expected to arise from two-loop diagrams in the UV completion of the pseudoscalar EFT. This is why we chose to normalise the $C_{a^2G}$ and $C_{a\tilde{G}}$ terms in Eq.~\eqref{eq:simpl-model-full} with a factor of $\alpha_s/(4\pi)$.

Let us first consider the one-loop contributions. The upper diagram in Fig.~\ref{fig:gluondiagrams} proceeds via the $a^2GG$ interaction and gives a contribution to the $\bar{\chi}\chi G^A_{\mu\nu}G^{A\mu\nu}$ operator:
\begin{align} \label{eq:CG0_a2GG}
    \mathcal{C}^{(0)}_G \supset -\frac{2}{9}\frac{C_\chi^2 C_{a^2G}}{\Lambda^2}\frac{m_\chi}{8\pi^2} F_0,
\end{align}
where, again, the loop function $F_0$ is given in App.~\ref{app:loopfunctions}.

The lower two diagrams in Fig.~\ref{fig:gluondiagrams} generate contributions to the $\bar{\chi}\chi G^A_{\mu\nu}G^{A\mu\nu}$ operator and the twist-2 gluon operators. (Similar to the analogous diagrams for the DM-quark operators.) The coefficients are
\begin{align}
    \mathcal{C}^{(0)}_G &\supset -\frac{2}{9}\frac{C_\chi^2 C_{a\tilde{G}}^2}{\Lambda^2} \frac{\alpha_s}{4\pi} \frac{m_\chi}{8\pi^2}(10 I_0 + 3m_\chi^2 I_1), \\[6pt]
    \mathcal{C}^{(1)}_G &= - \frac{C_\chi^2 C_{a\tilde{G}}^2}{\Lambda^2} \left(\frac{\alpha_s}{4\pi}\right)^2 \frac{I_0}{\pi^2},\\[6pt]
    \mathcal{C}^{(2)}_G &= - \frac{C_\chi^2 C_{a\tilde{G}}^2}{\Lambda^2} \left(\frac{\alpha_s}{4\pi}\right)^2 \frac{m_\chi I_1}{2\pi^2},
\end{align}
where the loop functions $I_n(m_\chi,m_a)$ are given in App.~\ref{app:loopfunctions}.

We now turn to the two-loop contributions mediated via the heavy quarks. Both the $a\bar{q} \gamma^5 q$ and $a^2\bar{q}q$ interactions induce the $\bar{\chi}\chi G^A_{\mu\nu}G^{A\mu\nu}$ interaction at two-loop order (see Fig.~\ref{fig:twoloop}).

To compute the upper diagram in Fig.~\ref{fig:twoloop}, which proceeds via the $a^2\bar{q}q$ interaction, we first integrate out the heavy quarks, leading to an effective $a^2 GG$ interaction. It is well known that to leading order in $1/m_q^2$, this corresponds to the substitution~\cite{Shifman:1978zn}
\begin{equation}
    m_q\bar{q}q \to-\frac{\alpha_s}{12\pi}G^{A}_{\mu\nu}G^{A\mu\nu}.
\end{equation}
Using this replacement, and computing the resulting one-loop diagram, we find that the contribution to the $\bar{\chi}\chi G^A_{\mu\nu}G^{A\mu\nu}$ operator is
\begin{equation}
    \mathcal{C}^{(0)}_G \supset \frac{2}{27}\sum_{q=c,b} \frac{C_\chi^2 C_{a^2q}}{\Lambda^2} \frac{m_\chi}{8\pi^2} F_0.
\end{equation}

The lower two diagrams in Fig.~\ref{fig:twoloop} proceed via the $a\bar{q} \gamma^5 q$ interaction. The lower diagram adds an additional complication, preventing one from straightforwardly integrating out the heavy quarks, as above, except in the limit $m_a \ll m_q$. This point has previously been discussed in detail in Refs.~\cite{Abe:2018emu,Ertas:2019dew}. The full two-loop calculation has been performed in Ref.~\cite{Abe:2018emu} and we use their result:
\begin{equation}
    \mathcal{C}^{(0)}_G \supset \sum_{q=c,b} -\frac{C_\chi^2 C_{aq}^2 m_\chi m_q^2}{432\pi^2\Lambda^2} F_3,
\end{equation}
where the loop function $F_3$ is equivalent to Eq.~(3.46) in Ref.~\cite{Abe:2018emu}.

\subsection{SI cross section}

Combining the above results, we obtain the general expressions for the couplings of the effective Lagrangian in Eq.~\eqref{eq:DM-quarkglu-effective} in terms of the Wilson coefficients of the pseudoscalar EFT in Eq.~\eqref{eq:simpl-model-full}. For the scalar operators, 
\begin{align}
    \begin{split}
    \mathcal{C}^{(0)}_q &= \frac{C_\chi^2 C_{a^2q}}{\Lambda^2} \frac{m_\chi}{8\pi^2} F_0 \\
    &\quad +\frac{C_\chi^2 C_{aq}^2}{\Lambda^2}\frac{m_q^2 m_\chi}{64\pi^2} (6F_1 + F_2),
    \end{split} \\
    \begin{split} \label{eq:xxgg-coeff}
    \mathcal{C}^{(0)}_G &= \frac{2}{27}\sum_{q=c,b} \frac{C_\chi^2 C_{a^2q}}{\Lambda^2} \frac{m_\chi}{8\pi^2} F_0\\
    &\quad -\sum_{q=c,b} \frac{C_\chi^2 C_{aq}^2}{\Lambda^2} \frac{m_\chi m_q^2}{432\pi^2} F_3\\
    &\quad -\frac{2}{9}\frac{C_\chi^2 C_{a\tilde{G}}^2}{\Lambda^2}\frac{\alpha_s}{4\pi}\frac{m_\chi}{8\pi^2}(10 I_0 + 3m_\chi^2 I_1)\\
    &\quad -\frac{2}{9}\frac{C_\chi^2 C_{a^2G}}{\Lambda^2}\frac{m_\chi}{8\pi^2} F_0,
    \end{split}
\end{align}
and the twist-2 operators
\begin{align}
    \mathcal{C}^{(1)}_q &= \frac{C_\chi^2 C_{aq}^2}{\Lambda^2} \frac{m_q^2}{8\pi^2} F_1,\\
    \mathcal{C}^{(2)}_q &= \frac{C_\chi^2 C_{aq}^2}{\Lambda^2} \frac{m_q^2}{16\pi^2 m_\chi} F_2,\\
    \mathcal{C}^{(1)}_G &= - \frac{C_\chi^2 C_{a\tilde{G}}^2}{\Lambda^2}\left(\frac{\alpha_s}{4\pi}\right)^2 \frac{1}{\pi^2} I_0,\\
    \mathcal{C}^{(2)}_G &= - \frac{C_\chi^2 C_{a\tilde{G}}^2}{\Lambda^2} \left(\frac{\alpha_s}{4\pi}\right)^2 \frac{m_\chi}{2\pi^2}  I_1.
\end{align}

To compute the SI DM-nucleon scattering cross section, we still need to evaluate the relevant nucleon matrix elements:
\begin{align}
    f_q^N &\equiv \frac{1}{m_N} \bra{N}m_q\bar{q}q\ket{N} , \\
    f_g^N &\equiv -\frac{9\alpha_s}{8\pi}\frac{1}{m_N}\bra{N}G^A_{\mu \nu}G^{A\mu \nu}\ket{N} ,
\end{align}
where $N=\{p,n\}$.\footnote{For simplicity, we neglect the small effect of isospin breaking and take $N=p$ throughout.} The gluon matrix element can be related to the quark matrix elements via the trace of the energy momentum tensor~\cite{Shifman:1978zn,Hill:2014yxa}:
\begin{equation}
    f_g^N = 1 - \sum_{q=u,d,s} f_q^N + \mathcal{O}(\alpha_s) .
\end{equation}
For the twist-2 operators, the hadronic matrix elements can be expressed in terms of parton distribution functions~\cite{Hill:2014yxa}
\begin{align}
    \begin{split}
    \bra{N}\mathcal{O}^{q}_{\mu\nu}\ket{N} &= \frac{1}{m_{N}} \left( p_{\mu} p_{\nu} - \frac{1}{4} m_{N}^{2} g_{\mu\nu} \right) \\ &\times \left[ q^{N}_{(2)} + \bar{q}^{N}_{(2)} \right],
    \end{split} \\
    \bra{N}\mathcal{O}^{G}_{\mu\nu}\ket{N}
    &= \frac{1}{m_{N}}
    \left( p_{\mu} p_{\nu} - \frac{1}{4} m_{N}^{2} g_{\mu\nu} \right) g^N_{(2)},
\end{align}
where $p$ is the momentum of the nucleon, and $q^N_{(2)}$, $\bar{q}^N_{(2)}$, and $g^N_{(2)}$ are second moments of the parton distribution functions.
The numerical values of $f_q^N$ and $f_g^N$ are given in App.~\ref{app:matrixelements}. We obtain the PDF moments by numerically integrating the central member of the NNPDF4.0 NNLO set \texttt{NNPDF40\_nnlo\_as\_01180}~\cite{NNPDF:2021njg}.

The SI DM-nucleon scattering cross section is then given by
\begin{equation}
    \sigma_\text{SI}=\frac{1}{\pi}\left(\frac{m_N m_\chi}{m_N+m_\chi}\right)^2|\lambda_\text{SI}|^2,
    \label{eq:SI-cross-section}
\end{equation}
with
\begin{equation}
    \begin{split}
        \lambda_{\rm SI} &= m_N\Bigg[\sum_{q=u,d,s}\mathcal{C}^{(0)}_q f_q^N + \mathcal{C}^{(0)}_G f^N_{g} \\
        &\quad + \frac{3}{4} \sum_{q \ne t}(m_\chi \mathcal{C}^{(1)}_q + m_\chi^2 \mathcal{C}^{(2)}_q)\\
        &\qquad \times [q^N_{(2)}+\bar{q}^N_{(2)}] \\
        &\quad + \frac{3}{4}\left(m_\chi \mathcal{C}^{(1)}_G + m_\chi^2 \mathcal{C}^{(2)}_G \right) g^N_{(2)} \Bigg].
        \label{eq:SI-amplitude}
    \end{split}
\end{equation}

\begin{figure}[t]
    \centering
    \includegraphics[width=\linewidth]{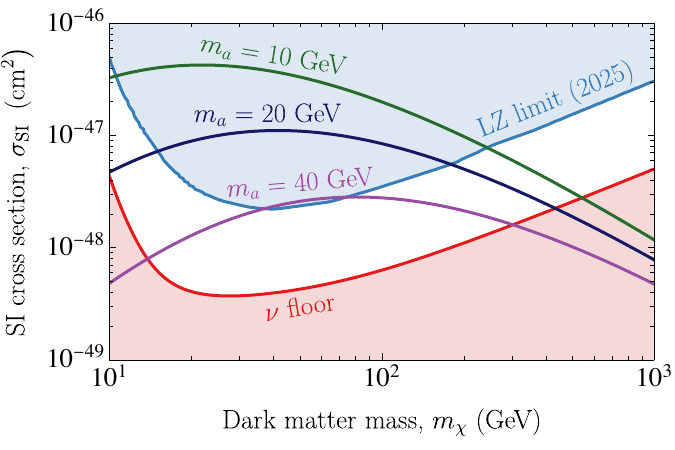}
    \caption{SI cross section induced by the $a\bar{\chi} \gamma^5 \chi$ and $a^2\bar{q}q$ interactions with $\Lambda=300~\mathrm{GeV}$ and $C_\chi=C_{a^2q}=1$ for $m_a=10~\mathrm{GeV}$ (green), $m_a=20~\mathrm{GeV}$ (blue), and $m_a=40~\mathrm{GeV}$ (purple). The blue shaded region is excluded by the LZ experiment~\cite{LZ:2024zvo}, while the red line shows the neutrino floor for xenon~\cite{OHare:2021utq}.}
    \label{fig:examplecs}
\end{figure}

An initial illustration of the magnitude of the loop-induced SI cross section is provided in Fig.~\ref{fig:examplecs}, for several benchmark values of $m_a$ with $\Lambda=300~\mathrm{GeV}$. This example assumes that only $C_\chi$ and $C_{a^2q}$ are non-zero (with $C_\chi = C_{a^2q} = 1$ and all other $C_i=0$). The latest exclusion limit from the LZ experiment~\cite{LZ:2024zvo} and the SI neutrino floor for $^{131}$Xe~\cite{OHare:2021utq} are shown for comparison, demonstrating that for reasonable values of the parameters, the SI cross section is in principle experimentally detectable.


\section{DM Annihilation and Constraints}
\label{sec:constraints}

We now discuss the constraints on the Wilson coefficients of the pseudoscalar EFT imposed by the DM relic abundance. We assume that the present-day DM abundance was produced through thermal freeze-out and require that the annihilation processes in the EFT yield the observed relic density.

\subsection{Annihilation processes}
\label{subsec:annihilationcs}

\begin{figure}[t]
    \centering
    \includegraphics{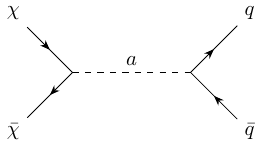}
    \includegraphics{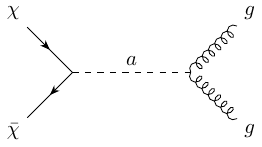}
    \includegraphics{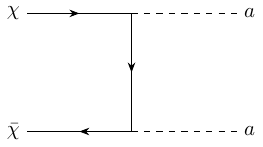}
    \includegraphics{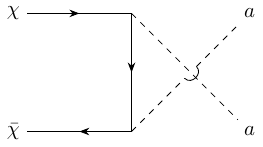}
    
    \caption{DM annihilation processes into pairs of quarks via the $a\bar{q} \gamma^5 q$ interaction (top), gluons via the $aG\tilde{G}$ interaction (second from top), and pseudoscalar mediators (bottom two).}
    \label{fig:dmannihilation}
\end{figure}

The tree-level, two-body annihilation channels are DM annihilation into pairs of quarks, gluons, or pseudoscalar mediators; the associated Feynman diagrams are depicted in Fig.~\ref{fig:dmannihilation}. We compute the thermally-averaged annihilation cross sections to leading non-zero order in relative velocity. 

The annihilation of DM into pseudoscalar pairs is kinematically allowed when $m_\chi > m_a$ and depends only on the $a\bar{\chi} \gamma^5 \chi$ coupling, $C_\chi$. It is therefore independent of the UV scale $\Lambda$. The thermally-averaged cross section is
\begin{equation}
    \label{eq:xxaa}
    \braket{\sigma v}_{\chi\bar{\chi} \to aa} \simeq C_\chi^4 \frac{m_\chi (m_\chi^2 - m_a^2)^{5/2}}{\pi (m_a^2 - 2 m_\chi^2)^4} \frac{1}{4x},
\end{equation}
where $x \equiv m_\chi/T$, with $T$ the temperature. 

The $a\bar{q} \gamma^5 q$ interaction allows for tree-level annihilation of DM into fermion pairs. For this annihilation channel,
\begin{equation}
    \label{eq:xxqq}
    \braket{\sigma v}_{\chi\bar{\chi} \to q\bar{q}} \simeq \sum_q \frac{C_\chi^2 C_{aq}^2}{\Lambda^2}\frac{N_c m_\chi^2 m_q^2}{2\pi (4 m_\chi^2 - m_a^2)^2} \sqrt{1-\frac{m_q^2}{m_\chi^2}},
\end{equation}
where $N_c=3$ and the sum extends over the quark flavours present in the EFT, with the $b\bar{b}$ channel providing the dominant contribution when kinematically accessible.

Finally, the $aG\tilde{G}$ interaction leads to annihilation into pairs of gluons, with
\begin{equation}
    \label{eq:xxgg}
    \braket{\sigma v}_{\chi\bar{\chi} \to gg} \simeq \frac{C_\chi^2 C_{a\tilde{G}}^2}{\Lambda^2}\frac{4 (N_c^2 -1) \left(\alpha_s/4\pi\right)^2 m_\chi^4}{\pi(4m_\chi^2 - m_a^2)^2}.
\end{equation}

In all of the above expressions, the Wilson coefficient $C_\chi$ is to be evaluated at the renormalisation scale $\mu=m_\chi$. In our numerical analysis in Sec.~\ref{sec:results}, we specify the Wilson coefficients at the scale $\mu=m_a$ and then account for the one-loop RGE running between $\mu=m_a$ and $\mu=m_\chi$ when computing the DM relic abundance and constraints from indirect detection. (This running is described in App.~\ref{app:running}.)

\subsection{Constraints}
\label{subsec:ann-constraints}

We constrain the Wilson coefficients by requiring that the available annihilation channels yield the observed DM relic abundance, $\Omega_\chi h^2 \simeq 0.12$~\cite{Planck:2018vyg}. We also impose constraints from indirect detection, where applicable. We distinguish two kinematic regimes:

\begin{enumerate}

\item $m_\chi>m_a$: In this regime the $\chi\bar{\chi}\to aa$ annihilation channel is open and typically dominates as the cross section is independent of the UV scale $\Lambda$ and is not suppressed by either $m_q$ or $\alpha_s/4\pi$, unlike annihilation into quarks or gluons, respectively. In this regime, the relic-density constraint can then be used to fix $C_\chi$ for given $m_\chi$ and $m_a$. Since the $\chi\bar{\chi}\to aa$ annihilation channel is $p$-wave, indirect detection does not provide any meaningful constraint.

\item $m_\chi<m_a$: The $\chi\bar{\chi}\to aa$ channel is kinematically forbidden, and annihilation proceeds only into SM final states. This occurs through the $\chi\bar{\chi}\to q\bar q$ or $\chi\bar{\chi}\to gg$ processes, provided that either $C_{aq}$ or $C_{a\tilde G}$ is non-zero. In the simple case that only one of these Wilson coefficients is non-zero, the relic-density constraint fixes the combination $C_\chi C_{aq}/\Lambda$ or $C_\chi C_{a\tilde G}/\Lambda$ for a given $m_\chi$ and $m_a$. In either case, the same combination of parameters also determines the loop-induced SI scattering rate.

The $\chi\bar{\chi}\to q\bar q$ and $\chi\bar{\chi}\to gg$ processes are $s$-wave and indirect detection can therefore constrain the same combination of parameters that determines the relic abundance (and the SI direct detection cross section). We impose the 95\% C.L. Fermi-LAT dwarf-spheroidal gamma-ray limits on annihilation into $b\bar b$ and $gg$ as recast in Ref.~\cite{Leane:2018kjk}, based on the \texttt{Pass 8} analyses of Refs.~\cite{Fermi-LAT:2015att,Fermi-LAT:2016uux}. These provide the strongest exclusion limits for the channels and parameter space we consider.

\end{enumerate}

Finally, note that in the region $m_\chi > \frac{3}{2}m_a$, the process $\chi\bar{\chi}\to aaa$ is kinematically accessible and also $s$-wave, for which reason one might expect it to be relevant both for determining the relic density and, in particular, for indirect detection. However, since this channel is both higher order in $C_\chi$ and suppressed by the 3-body phase space relative to $\chi\bar{\chi}\to aa$, it becomes relevant only when $C_\chi$ is close to unity. As we shall see in the next section, we find that the DM relic abundance condition requires $C_\chi\simeq 1$ only for $m_\chi\gtrsim 100$ GeV. In this mass regime, indirect detection is not currently sensitive to the thermal relic cross section. We therefore neglect the $\chi\bar{\chi}\to aaa$ channel in our analysis.


\section{Numerical Analysis}
\label{sec:results}

We now analyse the SI scattering signal produced by the pseudoscalar interactions in Eq.~\eqref{eq:simpl-model-full}, taking into account the constraints from relic density and indirect detection. For simplicity, we consider one non-zero SM--pseudoscalar coupling at a time, in combination with the DM--pseudoscalar coupling $C_\chi$. We explore the discoverability of each scenario---that is, its potential for generating observable SI scattering above the neutrino floor (as defined by Ref.~\cite{OHare:2021utq}) while remaining consistent with constraints from the relic abundance and indirect detection. We take this as the limit of discoverability in the near term, as reaching the neutrino floor is the goal of proposed future direct-detection experiments like XLZD~\cite{XLZD:2024nsu}. More sophisticated DM-discriminating strategies, such as directional detection, may eventually improve this limit, but we will not consider them here.

We consider the pseudoscalar mass range $m_a \in [5,100]~\mathrm{GeV}$, with the lower bound chosen to avoid otherwise stringent bounds on the pseudoscalar mediator from meson decays~\cite{Arcadi:2017wqi}. For the DM mass, we consider the range from 5\,GeV to the electroweak scale, with the upper bound determined by the fact that our EFT is defined below the electroweak scale. The lower bound is chosen such that freeze-out occurs prior to the quark-hadron phase transition at $T_C \sim 151\,\mathrm{MeV}$~\cite{Aoki:2009sc} (given that the freeze-out temperature is $T_f\simeq m_\chi/30\gtrsim 160$\,MeV for $m_\chi\gtrsim 5$\,GeV).

\subsection{\texorpdfstring{$a^2\bar{q}q$}{a2qq} scenario}

\begin{figure}[t]
    \centering
    \includegraphics[width=\linewidth]{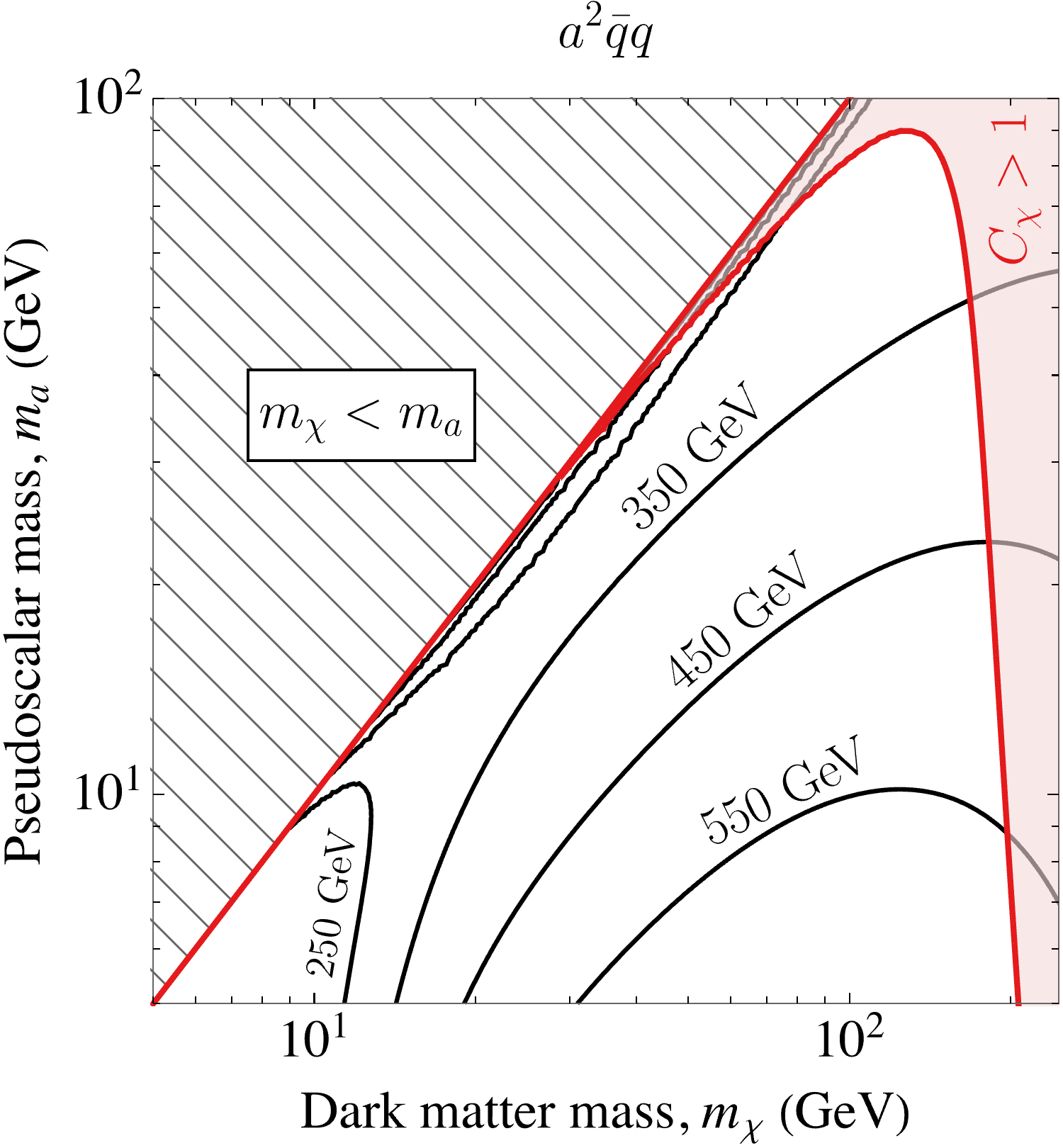}
    \caption{The $m_\chi$\,--\,$m_a$ parameter space in the $a^2\bar{q}q$ scenario for $m_\chi>m_a$. Here, the relic density is set by the $\chi\bar{\chi} \to aa$ process. The contours show the values of $\Lambda$ which correspond to a SI cross section at the neutrino floor. The red shaded region is where the relic density requires $C_\chi>1$ and the EFT is no longer perturbative.}
    \label{fig:aaqqlower}
\end{figure}

We begin with the case where the pseudoscalar couples to the SM via $a^2\bar{q}q$. In this scenario, the only non-zero Wilson coefficients are $C_\chi$ and $C_{a^2q}$, and we set the latter to unity without loss of generality. The relevant parameters are therefore $\{m_\chi, m_a, C_\chi, \Lambda\}$. Fig.~\ref{fig:aaqqlower} shows our results in the $(m_\chi, m_a)$ plane; we partition the space by the kinematic threshold $m_\chi=m_a$ and treat the two regimes separately.

When \mbox{$m_\chi > m_a$}, the $\chi\bar{\chi} \to aa$ annihilation channel is open and the relic abundance constraint fixes $C_\chi$ at each point in the $m_\chi$\,--\,$m_a$ parameter space, as outlined in Sec.~\ref{subsec:ann-constraints}; the SI cross section is then determined solely by the scale $\Lambda$. The black contours in Fig.~\ref{fig:aaqqlower} show the values of $\Lambda$ that correspond to a SI cross section at the neutrino floor. We also flag the region where $C_\chi > 1$ (red shaded region), in which it is no longer possible to attain the correct relic abundance with perturbative couplings.

In the \mbox{$m_\chi > m_a$} region, values of $\Lambda$ up to $\sim 640~\mathrm{GeV}$ can yield a SI cross section at or above the neutrino floor. This is the largest value of $\Lambda$ obtained in any of the specific scenarios we have studied. Well-motivated UV completions that realise this scenario involve extended scalar sectors, such as two-Higgs doublet models with an additional pseudoscalar state~\cite{Arcadi:2017wqi, Bauer:2017ota, Abe:2018emu, Arcadi:2022lpp, Arcadi:2026aau}. In these theories, the effective $a^2\bar{q}q$ interaction is generated via the operator $a^2 H^\dagger H$ after electroweak symmetry breaking.

When $m_\chi < m_a$, there are no tree-level annihilation processes when $a^2\bar{q}q$ is the only non-zero SM--pseudoscalar coupling. The leading annihilation channel is to quarks via the one-loop diagram at the top of Fig.~\ref{fig:quarkdiags} (rotated ninety degrees), which has a cross section proportional to $(C_\chi^2 C_{a^2q} m_q/(16\pi^2\Lambda^2))^2$. Even when setting the relevant $C_i=1$ (the most optimistic limit), a naive calculation suggests that $\Lambda \sim \mathcal{O}(1)~\mathrm{GeV}$ is required to reproduce the observed relic abundance. The UV-completion must therefore contain light new states coupled to quarks and we clearly cannot compute the relic abundance in this region of parameter space without including these states in the EFT. Furthermore, any such UV completion is likely to be strongly constrained by precision observables. For these reasons we will not study the $m_\chi<m_a$ region in detail.

Note that when $a^2\bar{q}q$ is the only non-zero coupling to the SM, the pseudoscalar mediator has no decay channels at tree level; however, no symmetry forbids the decay of the mediator at loop level. In the event that the rate of this decay is suppressed such that the pseudoscalar is long-lived---thereby potentially coming into tension with constraints from Big Bang nucleosynthesis---even a small but non-zero $a\bar{q}\gamma^5 q$ coupling of order $C_{aq}\gtrsim\mathcal{O}(10^{-7})$ will suffice to ensure that $a$ decays in less than a second. Clearly, the direct-detection phenomenology of this scenario will not be meaningfully impacted.

\subsection{\texorpdfstring{$a\bar{q} \gamma^5 q$}{aqq} scenario}

\begin{figure}[t]
    \centering
    \includegraphics[width=\linewidth]{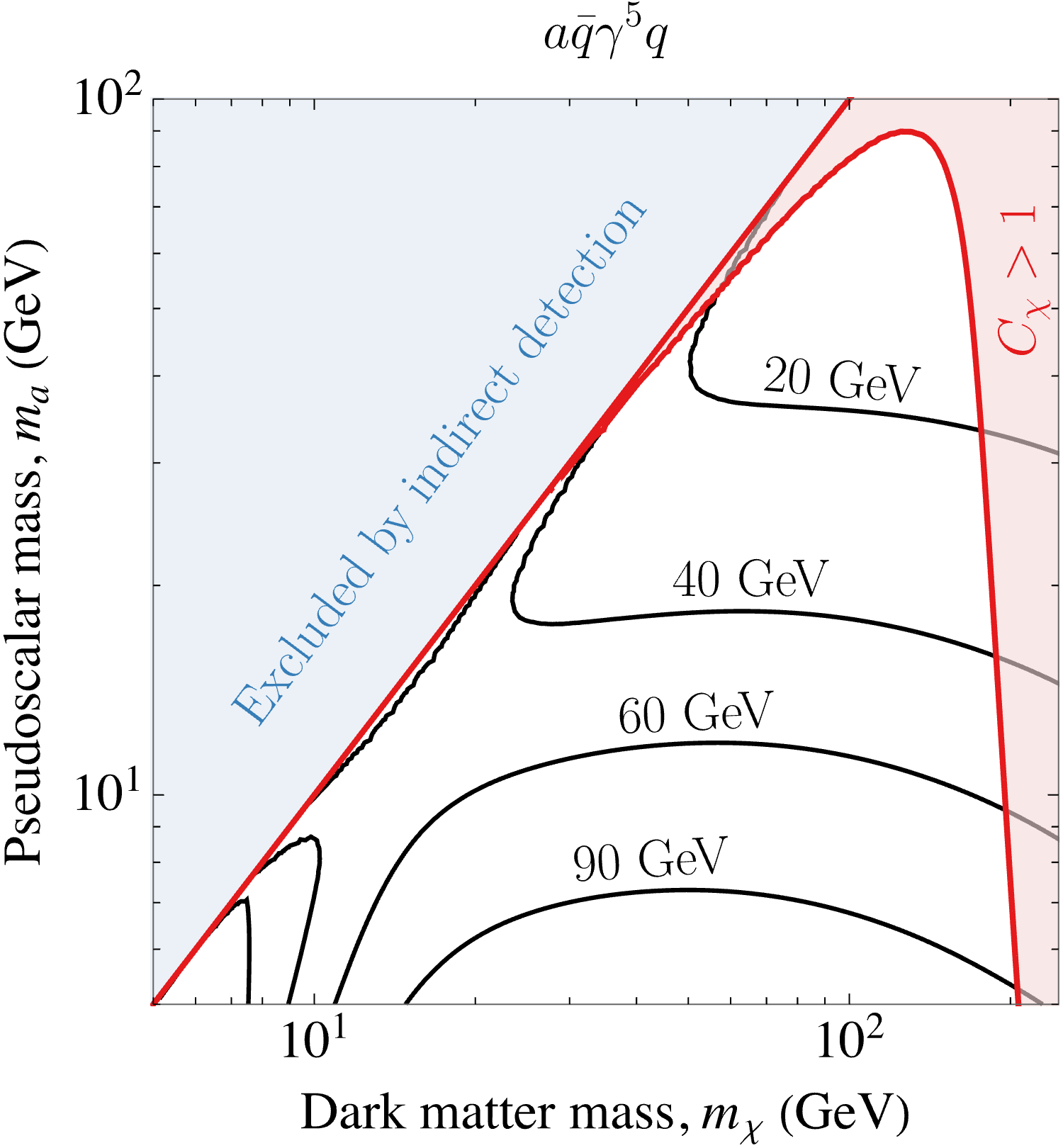}
    \caption{The $m_\chi$\,--\,$m_a$ parameter space in the $a\bar{q} \gamma^5 q$ scenario. In the $m_\chi<m_a$ region, the relic density is set by the $\chi\bar{\chi} \to q\bar{q}$ process; this region is excluded by the constraints from Fermi-LAT~\cite{Leane:2018kjk}, as indicated by the blue shading. In the $m_\chi > m_a$ region, the contours show the values of $\Lambda$ which correspond to a SI cross section at the neutrino floor. The red shaded region is where the relic density requires $C_\chi>1$ and the EFT is no longer perturbative.}
    \label{fig:aqq}
\end{figure}

Next, we consider the case where the pseudoscalar couples to the SM via $a\bar{q}\gamma^5 q$, with the $m_\chi$\,--\,$m_a$ parameter space shown in Fig.~\ref{fig:aqq}. Again, we partition the space by the kinematic threshold $m_\chi=m_a$. 

In the $m_\chi<m_a$ region, DM annihilation is governed by the $s$-wave channel $\chi\bar{\chi}\to q\bar{q}$. We then find that the indirect-detection constraint from the Fermi-LAT dwarf analysis already excludes the entire range of DM masses we consider (for $m_\chi<m_a$), as shown by the blue shaded region in Fig.~\ref{fig:aqq}. Note that even allowing for the possibility that the indirect-detection constraint may be relaxed due to uncertainties in the astrophysical $J$-factors, we find that the SI scattering cross section would lie below the neutrino floor throughout the vast majority of the $m_\chi<m_a$ region.

On the other hand, in the $m_\chi>m_a$ region, the $p$-wave annihilation process $\chi\bar{\chi}\to aa$ dominates and the relevant features are determined analogously to the $a^2\bar{q}q$ scenario detailed above. The difference is that, here, the values of $\Lambda$ required to obtain a SI cross section at or above the neutrino floor are restricted to $\Lambda \lesssim 120~\mathrm{GeV}$, and new particles with mass below the electroweak scale would be required to UV-complete the theory. Such particles may already be subject to constraints from direct searches at colliders. Furthermore, note that for a good fraction of this parameter space, the DM mass $m_\chi$ exceeds the values of $\Lambda$ required for a direct-detection signal; here, the DM abundance cannot be reliably computed within the EFT. In particular, the new light states could enable new DM annihilation channels. As such, the $\Lambda$ contours should be understood only as an indication that observable SI scattering cannot occur in this parameter space without light new physics.

\subsection{\texorpdfstring{$a^2GG$}{a2GG} scenario}

\begin{figure}[t]
    \centering
    \includegraphics[width=\linewidth]{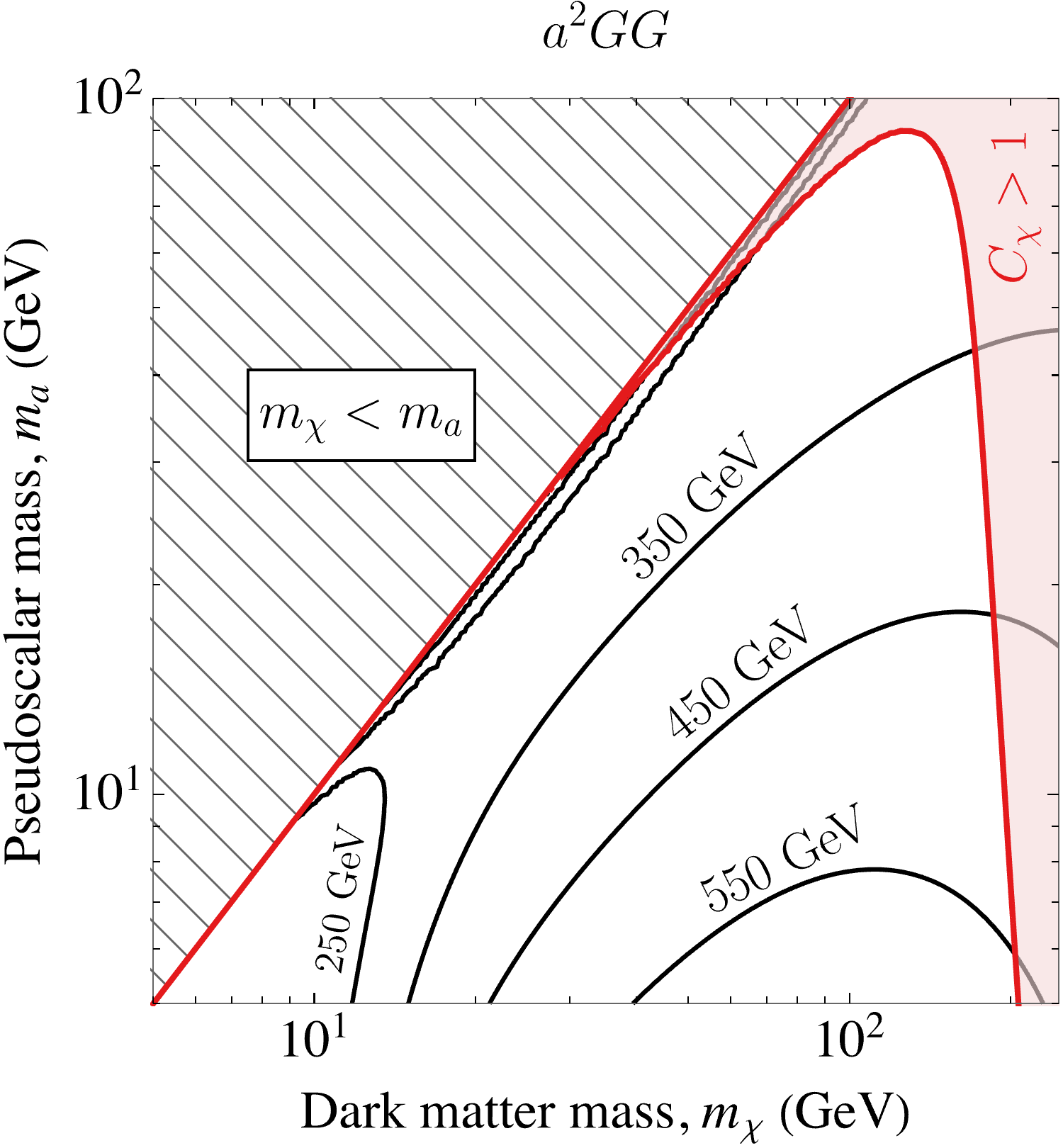}
    \caption{The $m_\chi$\,--\,$m_a$ parameter space in the $a^2GG$ scenario for $m_\chi>m_a$. Here, the relic density is set by the $\chi\bar{\chi} \to aa$ process. The contours show the values of $\Lambda$ which correspond to a SI cross section at the neutrino floor. The red shaded region is where the relic density requires $C_\chi>1$ and the EFT is no longer perturbative.}
    \label{fig:aaGG}
\end{figure}

Figure~\ref{fig:aaGG} shows the results for the case where the pseudoscalar couples to the SM via $a^2 GG$. As in the $a^2\bar{q}q$ case, we do not consider the $m_\chi<m_a$ region in detail; here, the leading annihilation channel, $\chi\bar\chi \to gg$, arises at loop level and, consequently, very low values of $\Lambda$ are required to achieve the correct relic abundance. To reliably explore this region of parameter space, one must therefore include these new light degrees of freedom within the EFT.

On the other hand, in the $m_\chi>m_a$ region, the values of $\Lambda$ that yield a SI cross section at the neutrino floor are broadly similar to those in the $a^2 \bar{q}q$ scenario, albeit with a slightly lower maximum of $\Lambda \simeq 605~\mathrm{GeV}$ (where we have taken $C_{a^2G}=1$ without loss of generality). The origin of this similarity can be immediately seen from the expressions in Eqs.~\eqref{eq:Cq0_a2qq} and \eqref{eq:CG0_a2GG}, where the SI amplitude is proportional to the same loop function, $F_0(m_\chi,m_a)$, in both cases.

Similar to the $a^2\bar{q}q$ scenario, when $a^2 GG$ is the only coupling of the pseudoscalar to the SM, the pseudoscalar has no tree-level decays. However, even a tiny $aG\tilde{G}$ coupling is sufficient to allow $a$ to efficiently decay, while having negligible effect on the relic density or SI cross section. (In fact, such a coupling is expected in realistic UV completions, as discussed below.)

The $a^2GG$ scenario is readily UV-completed via either the top quark (assuming that it couples to $a$) or new heavy coloured fermions that couple to the pseudoscalar~\cite{Fan:2015sza}. The former case has the advantage of minimality---it requires no heavy degrees of freedom outside the SM---and, since $\Lambda>m_t$ throughout the $m_\chi>m_a$ region, we expect that an $\mathcal{O}(1)$ pseudoscalar-top coupling would generate a SI signal that is comfortably visible to experiment. On the other hand, UV completions with new heavy coloured fermions may be in tension with direct searches at the LHC (see e.g. Ref.~\cite{ATLAS:2018mpo}). Note that while these completions would induce both the $a^2 GG$ and $aG\tilde{G}$ operators within the EFT, the direct-detection phenomenology in such scenarios is likely dominated by the $a^2GG$ operator. The reason for this can be seen in Eq.~\eqref{eq:xxgg-coeff}, where the contribution from $C_{a\tilde{G}}$ to $\mathcal{C}_G^{(0)}$ is suppressed by an additional power of $\alpha_s/4\pi$.

\subsection{\texorpdfstring{$aG\tilde{G}$}{aGG\~} scenario}

\begin{figure}[t]
    \centering
    \includegraphics[width=\linewidth]{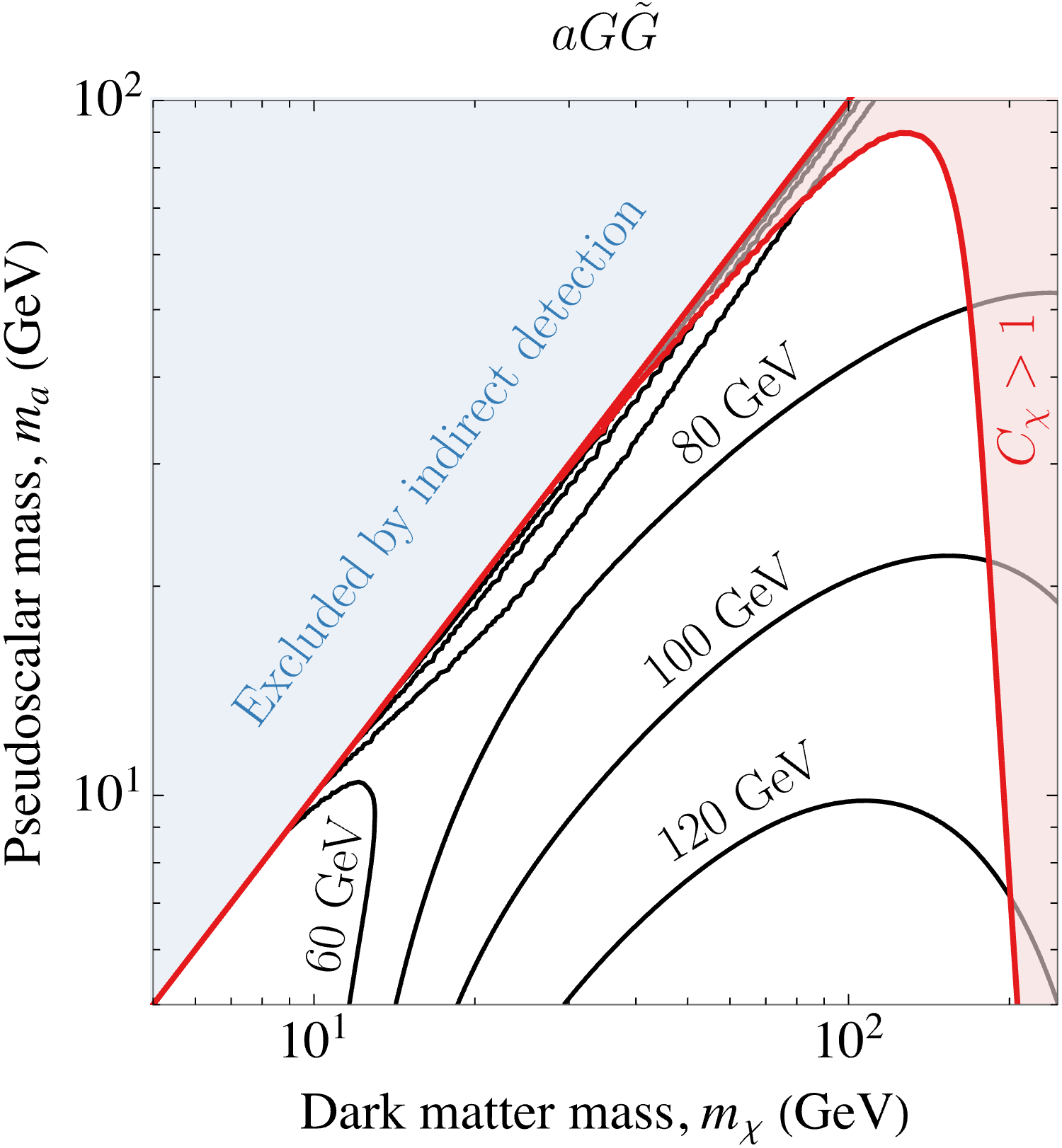}
    \caption{The $m_\chi$\,--\,$m_a$ parameter space in the $aG\tilde{G}$ scenario. In the $m_\chi<m_a$ region, the relic density is set by the $\chi\bar{\chi} \to gg$ process; this region is excluded by the constraints from Fermi-LAT~\cite{Leane:2018kjk}, as indicated by the blue shading. In the $m_\chi>m_a$ region, the contours show the values of $\Lambda$ which correspond to a SI cross section at the neutrino floor. The red shaded region is where the relic density requires $C_\chi>1$ and the EFT is no longer perturbative.}
    \label{fig:aGG}
\end{figure}

Figure~\ref{fig:aGG} shows the $m_\chi$\,--\,$m_a$ parameter space in the scenario where the pseudoscalar couples to the SM via $aG\tilde{G}$. In the $m_\chi<m_a$ region, DM annihilation is governed by the $s$-wave channel $\chi\bar{\chi}\to gg$. As such, we find that indirect-detection constraints already exclude this region of parameter space, as shown by the blue shaded region in Fig.~\ref{fig:aGG}. Temporarily putting this constraint aside, we also find that SI scattering would be undetectable for the majority of the $m_\chi<m_a$ region.

In the $m_\chi > m_a$ regime, the values of $\Lambda$ that generate an observable SI signal are restricted to $\lesssim 140$\,GeV, corresponding to new physics at or below the electroweak scale which may be subject to bounds from direct searches. Note that here, unlike in the $a\bar{q}\gamma^5 q$ scenario, the values of $\Lambda$ that lead to an observable SI signal are significantly greater than the DM mass $m_\chi$ across most of the parameter space and the relic density can be reliably computed within the EFT. 

As previously mentioned, minimal UV completions of the $aG\tilde{G}$ operator involving heavy coloured fermions will also generate the $a^2GG$ operator. Such scenarios include the minimal top quark UV completion. As discussed in the previous section, in any UV completion that generates comparable Wilson coefficients for these operators, the direct-detection phenomenology is expected to be dominated by the $a^2GG$ operator.


\section{Conclusions}
\label{sec:conclusion}

In this work, we revisited the direct-detection prospects for Dirac-fermion DM that interacts with the SM via a purely pseudoscalar portal at tree level, working in the model-agnostic context of a sub-electroweak EFT containing the pseudoscalar mediator. We considered operators up to dimension-six, imposing minimal flavour violation. The primary goal was to determine whether loop-induced spin-independent scattering can generate an experimentally relevant signal within this general scenario.

We computed the loop-induced spin-independent DM--nucleon scattering generated by each operator structure in the EFT. We then considered several benchmark scenarios to assess whether the resulting direct-detection signal can lie above the neutrino background, while simultaneously satisfying the relic-density constraint and bounds from indirect detection.

For operators that are quadratic in $a$ (i.e. $a^2 m_q\bar{q}q/\Lambda^2$ and $a^2GG/\Lambda^2$), we find that when $m_\chi > m_a$ there are significant regions of parameter space in which the SI cross section lies above the neutrino floor. These correspond to $\Lambda\lesssim640$\,GeV, with the largest values of $\Lambda$ achieved for a light pseudoscalar mediator, $m_a\simeq 5\,\mathrm{GeV}$. Direct detection may therefore provide a means to probe such scenarios in the future. 

On the other hand, the operators linear in $a$ (i.e. $m_q a\bar{q} \gamma^5 q/\Lambda$ and $aG\tilde{G}/\Lambda$) yield a SI signal above the neutrino floor only when the scale $\Lambda$ lies at or below the electroweak scale. In this case, new light particles must be present in the spectrum of the underlying high-energy theory, and direct searches for these states likely already constrain such scenarios. 

Although we have considered one pseudoscalar--SM operator at a time, we expect the $a^2\bar{q}q$ and $a^2GG$ interactions to generally dominate the SI amplitude when several operators are present.

Therefore, though it remains the case that pseudoscalar-mediated dark matter is in many cases invisible to direct detection (while remaining a perfectly viable model of dark matter), there are scenarios in which the loop-induced SI scattering may yield a visible signal. For this signal to be observable at the next generation of experiments requires additional new physics well below the TeV scale. This situation can arise, for example, in UV completions with extended Higgs sectors~\cite{Arcadi:2017wqi, Bauer:2017ota, Abe:2018emu, Arcadi:2022lpp}, and it would be interesting to explore this in the context of other UV completions for pseudoscalar-mediated dark matter.

\begin{acknowledgments}
This work was supported in part by the ARC Centre of Excellence for Dark Matter Particle Physics, CE200100008. L.W.F.\ is supported by a Melbourne Research Scholarship. E.N.V.W.\ is supported by an Australian Government Research Training Program Scholarship and the Elizabeth and Vernon Puzey Scholarship. Feynman diagrams were generated using the Ti\textit{k}Z-Feynman package for \LaTeX~\cite{Ellis:2016jkw}. 
\end{acknowledgments}

\appendix


\section{Pseudoscalar contact interactions}
\label{app:contact}

In the EFT defined in Sec.~\ref{sec:eft-setup} and used throughout the paper, we retain the pseudoscalar mediator as a low-energy degree of freedom in the effective theory. Here, we explain this choice by comparing with an EFT that contains only the pseudoscalar contact interactions that would be obtained by integrating out $a$ at tree level. The important point is that such an EFT does not capture the leading SI contributions generated by a UV model, which arise from one-loop matching. Subject to the field-content, symmetry, and power-counting assumptions of Sec.~\ref{sec:eft-setup}, retaining $a$ yields the most general sub-electroweak theory in which these threshold contributions are calculable.

The leading pseudoscalar contact interactions are
\begin{align}
\label{eq:pure-p-contact}
    \mathcal{L}_\text{int}&=\sum_{q\neq t}\frac{C_{q}^P}{\Lambda^{3}} m_q \bar{\chi}i\gamma^5\chi\bar{q}i\gamma^5 q+\frac{C_{G}^P}{\Lambda^{3}}\bar{\chi}i\gamma^5\chi G^{A}_{\mu\nu}\tilde{G}^{A\mu\nu} \ ,
\end{align}
where $\Lambda$ here is a bookkeeping scale for the contact EFT and should not be identified with the $\Lambda$ of Eq.~\eqref{eq:simpl-model-full}.

\begin{figure}[t]
    \centering
    \includegraphics{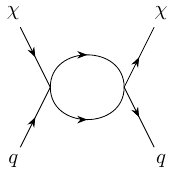}
    \hspace{1em}
    \includegraphics{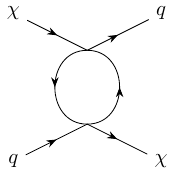}
    \caption{Diagrams generating the four-fermion scalar interactions through one-loop running assuming only DM--quark pseudoscalar contact interactions. There are analogous diagrams for pseudoscalar DM--gluon interactions.}
    \label{fig:heavy-med-loops}
\end{figure}

We have imposed $C_i^{\rm SI}(\mu_0)=0$, where the $C_i^{\rm SI}$ represent the Wilson coefficients of the operators that generate unsuppressed SI scattering, and $\mu_0$ is the reference scale at which the conditions hold. We highlight that this pseudoscalar-only boundary condition is not radiatively stable. Two insertions of the pseudoscalar-current operators of Eq.~\eqref{eq:pure-p-contact} will generate scalar-current operators at lower scales through one-loop diagrams such as those shown in Fig.~\ref{fig:heavy-med-loops}. Focusing on the example of the quark diagrams, and defining the scalar operator as
\begin{equation}
\mathcal L \supset
\frac{C_{q}^{S}}{\Lambda^3}
m_q\bar\chi\chi\bar q q,
\end{equation}
the generated scalar coefficient is schematically
\begin{equation}
\label{eq:running-cs}
\frac{m_q C_{q}^{S}(\mu)}{\Lambda^3} \sim
\frac{(C_{q}^{P})^2}{16\pi^2}
\frac{m_\chi m_q^3}{\Lambda^6}
\log \frac{\mu_0}{\mu}.
\end{equation}

This should be compared with the generic matching contribution to the same scalar operator. In the absence of a symmetry or tuning that forbids it, integrating out UV physics at the scale $\mu_0$ will generate $C_{q}^S$. This contribution enters the direct-detection amplitude proportional to $m_q/\Lambda^3$, rather than $m_\chi m_q^3/\Lambda^6$ as in Eq.~\eqref{eq:running-cs}. Thus, for comparable dimensionless coefficients, the matching contribution dominates parametrically over the contribution generated by the RG running.

While Eq.~\eqref{eq:running-cs} contributes to a (tiny) radiative SI cross section, we are really interested in calculating the dominant contribution to $C_{q}^{S}$ coming from matching. We therefore consider an EFT that includes the pseudoscalar mediator $a$ explicitly, and then integrate it out at one loop to compute the resulting direct detection cross section. This is the approach used in the main text of the paper.

\section{Renormalisation-group evolution and operator mixing}
\label{app:running}

In this appendix, we summarise the renormalisation-group evolution of the couplings and Wilson coefficients appearing in Eq.~\eqref{eq:simpl-model-full}.

First, consider the coupling $C_\chi$ which controls the strength of the interaction between the DM and the mediator. The one-loop renormalisation group equation for $C_\chi$ is~\cite{Ghorbani:2017qwf}
\begin{equation}
    \mu\frac{dC_\chi}{d\mu}
    =\frac{5}{16\pi^2}C_\chi^3,
    \label{eq:Cchi-running}
\end{equation}
with solution
\begin{equation}
    C_\chi(\mu)
    =
    \frac{C_\chi(\mu_0)}
    {\sqrt{1-\dfrac{5C_\chi^2(\mu_0)}{8\pi^2}
    \log\left(\dfrac{\mu}{\mu_0}\right)}},
    \label{eq:Cchi-running-solution}
\end{equation}
such that $C_\chi$ increases towards higher scales. 

Renormalisation-group evolution also mixes a subset of the operators in Eq.~\eqref{eq:simpl-model-full}. Since $\chi$ and $a$ are singlets under the SM gauge group, this mixing is determined by the SM bilinears, for which the relevant anomalous dimensions are well known. The $CP$-even SM operators $m_q\bar{q}q$ and $\alpha_s GG$ are scale-invariant and do not mix at leading order in $\alpha_s$, and thus the coefficients, $C_{a^2q}$ and $C_{a^2G}$, are approximately RG-invariant~\cite{Hisano:2015bma, Grinstein:1988wz}. By contrast, the $CP$-odd coefficients $C_{aq}$ and $C_{a\tilde{G}}$ mix under RG running. As a function of initial scale $\mu_0$ and final scale $\mu$, the coefficients are related by~\cite{Bauer:2020jbp}
\begin{equation}
    C_{aq}(\mu)=C_{aq}(\mu_0)+\frac{4}{\beta_0}\frac{\alpha_s(\mu)-\alpha_s(\mu_0)}{\pi}\Sigma(\mu_0),
\end{equation}
and
\begin{equation}
    C_{a\tilde{G}}(\mu)=C_{a\tilde{G}}(\mu_0)+\frac{2n_f}{\beta_0}\frac{\alpha_s(\mu)-\alpha_s(\mu_0)}{\pi}\Sigma(\mu_0),
\end{equation}
where $\beta_0=11-\tfrac{2}{3}n_f$, with $n_f$ the number of active quark flavours, and we define
\begin{align}
    \Sigma(\mu_0)&\equiv C_{a\tilde{G}}(\mu_0)-\frac{n_f}{2} C_{aq}(\mu_0).
\end{align}


\section{Loop functions}
\label{app:loopfunctions}

We collate the loop functions relevant to Sec.~\ref{sec:SI-scattering} below. These expansions were evaluated with \texttt{Package-X}~\cite{Patel:2016fam}. The loop function associated with the $a^2\bar{q}q$ and $a^2 GG$ interactions is given by
\begin{align}
    F_0(m_\chi,m_a) &= -\frac{1}{m_\chi^2} - \frac{\left(m_\chi^2 -m_a^2\right)}{2m_\chi^4}\log\left(\frac{m_a^2}{m_\chi^2}\right) \nonumber \\[6pt]
    &\quad - \frac{m_a\left(m_a^2 - 3m_\chi^2 \right)}{m_\chi^4 \sqrt{m_a^2 - 4m_\chi^2}} \nonumber\\[6pt]
    &\qquad \times \log{\left(\frac{m_a^2 + \sqrt{m_a^2 \left(m_a^2 - 4m_\chi^2 \right)}}{2m_a m_\chi}\right)}.
\end{align}
In the limit of zero quark mass, the loop functions associated with the $a\bar{q}\gamma^5q$ interactions are given by
\begin{align}
    F_1(m_\chi,m_a) &= -\frac{2m_a^2-3m_\chi^2}{6m_a^2 m_\chi^4} +\frac{m_a^2-3m_\chi^2}{6m_\chi^6}\log{\left(\frac{m_a^2}{m_\chi^2}\right)} \nonumber \\[6pt] 
    &\quad + \frac{\sqrt{m_a^2(m_a^2-4m_\chi^2)}(m_\chi^2-m_a^2)}{3m_a^2 m_\chi^6} \nonumber \\[6pt]
    &\qquad \times \log{\left(\frac{m_a^2+\sqrt{m_a^2(m_a^2-4m_\chi^2)}}{2m_a m_\chi}\right)},
\end{align}
\begin{align}
    F_2(m_\chi,m_a) &= \frac{2m_a^2-m_\chi^2}{m_a^2 m_\chi^4}-\frac{m_a^2-2m_\chi^2}{m_\chi^6}\log{\left(\frac{m_a^2}{m_\chi^2}\right)} \nonumber \\[6pt]
    &\quad + \frac{2(m_a^4-4m_a^2 m_\chi^2+2m_\chi^4)}{m_a m_\chi^6\sqrt{m_a^2-4m_\chi^2}} \nonumber \\[6pt]
    &\qquad \times \log{\left(\frac{m_a^2+\sqrt{m_a^2(m_a^2-4m_\chi^2)}}{2m_a m_\chi}\right)}.
\end{align}
Note that we retain the full $m_q$ dependence in our analysis, but this is typically a small correction to the zeroth-order expressions listed above. 

Finally, the loop functions associated with the $aG\tilde{G}$ interactions are given by
\begin{align}
    I_0(m_\chi,m_a) &= -\frac{5m_\chi^2-2m_a^2}{8m_\chi^4} \nonumber \\
    &\quad -\frac{2m_\chi^4-4m_\chi^2 m_a^2+m_a^4}{8m_\chi^6}\log{\left(\frac{m_a^2}{m_\chi^2}\right)} \nonumber \\[6pt]
    &\quad -\frac{\sqrt{m_a^2(m_a^2-4m_\chi^2)}(2m_\chi^2-m_a^2)}{4m_\chi^6} \nonumber \\[6pt]
    &\qquad \times\log{\left(\frac{m_a^2+\sqrt{m_a^2(m_a^2-4m_\chi^2)}}{2m_\chi m_a}\right)},
\end{align}
\begin{align}
    I_1(m_\chi,m_a) &=\frac{7m_\chi^2-6m_a^2}{4m_\chi^6} \nonumber \\
    &\quad +\frac{2m_\chi^4-8m_\chi^2 m_a^2+3m_a^4}{4m_\chi^8}\log{\left(\frac{m_a^2}{m_\chi^2}\right)} \nonumber \\[6pt]
    &\quad -\frac{m_a(12m_\chi^4-14m_\chi^2 m_a^2+3m_a^4)}{2m_\chi^8\sqrt{m_a^2-4m_\chi^2}} \nonumber \\
    &\qquad \times \log{\left(\frac{m_a^2+\sqrt{m_a^2(m_a^2-4m_\chi^2)}}{2m_\chi m_a}\right)}.
\end{align}


\section{Hadronic matrix elements}
\label{app:matrixelements}

We use the values of the nucleon matrix elements from Ref.~\cite{FLAG:2024oxs}:
\begin{equation}
\begin{split}
    f_u^p &= 0.0164 \pm 0.0011, \\
    f_d^p &= 0.0283 \pm 0.0019, \\
    f_s^p &= 0.0479 \pm 0.0034.
\end{split}
\end{equation}
At leading order in $\alpha_s$, the gluon matrix element is
\begin{equation}
    f_g^p = 1 - \sum_{q=u,d,s} f_q^p = 0.9075 \pm 0.0040 .
\end{equation}

We evaluate the twist-2 matrix elements and their corresponding Wilson coefficients at the factorisation scale $\mu=m_a$. For both the nucleon matrix elements and the second moments of the PDFs extracted from Ref.~\cite{NNPDF:2021njg}, we work with the central values and neglect the corresponding uncertainties.


\newpage
\bibliographystyle{apsrev4-2}
\bibliography{main}

\end{document}